# Exploring the Creation and Humanization of Digital Life: Consciousness Simulation and Human-Machine Interaction

Qikang Zhang
South China Normal University

*Abstract*—Digital life, a form of life generated by computer programs or artificial intelligence systems, it possesses self-awareness, thinking abilities, emotions, and subjective consciousness. Achieving it involves complex neural networks, multi-modal sensory integration [1, 2], feedback mechanisms, and self-referential processing [3]. Injecting prior knowledge into digital life structures is a critical step. It guides digital entities' understanding of the world, decision-making, and interactions. We can customize and personalize digital life, it includes adjusting intelligence levels, character settings, personality traits, and behavioral characteristics. Virtual environments facilitate efficient and controlled development, allowing user interaction, observation, and active participation in digital life's growth. Researchers benefit from controlled experiments, driving technological advancements. The fusion of digital life into the real world offers exciting possibilities for human-digital entity collaboration and coexistence.

## 1 INTRODUCTION

Recent years have witnessed a remarkable convergence of knowledge from diverse fields, leading to the emergence of a novel frontier: conscious digital life. This nascent domain seeks to imbue computer programs and AI systems with attributes resembling consciousness, including self-awareness, emotional experiences, and cognitive abilities.

The significance of conscious digital life transcends mere technological innovation. It ventures into the realms of philosophy, ethics, and human-computer interaction. If successful, this endeavor could reshape our understanding of what it means to be conscious and potentially challenge existing philosophical frameworks. Moreover, it has profound implications for fields ranging from AI ethics to the creation of sentient machines that coexist with humans.

Previous attempts to replicate consciousness in artificial systems have primarily focused on narrow aspects of cognition or specific tasks. These efforts, while commendable, have fallen short of simulating the holistic nature of human consciousness. The distinction in our approach lies in the ambition to create digital entities that encompass the multifaceted qualities of consciousness, from sensory integration to emotional experiences, in a unified manner.

This paper embarks on a journey to explore the creation and development of conscious digital life, aiming to elucidate the methods, challenges, and potential implications of this groundbreaking endeavor. In the following sections, we delve deeper into these facets, providing insights into the intricate journey of creating conscious digital life and the potential consequences it holds for the future of technology, science, and our understanding of consciousness itself.

## 2 THE EMERGENCE OF CONSCIOUSNESS

### 2.1 DEFINITION AND COMPLEXITY OF CONSCIOUSNESS

Consciousness is a complex and multifaceted concept with varying definitions stemming from different disciplines and perspectives. It can be regarded as a synthesis of subjective experiences, perception, thinking, self-awareness, and emotions. The nature and characteristics of consciousness [4] vary based on research within fields such as philosophy, neuroscience, and psychology. Our discussion primarily unfolds from a bionics perspective.

The emergence of consciousness is purposeless and not reliant on specific brain structures. Instead, it can be viewed as an expression of a neural system with a certain level of complexity, harboring a non-conceptual theory of self. This theory is formed through experiences gained from interactions with oneself, the external world, and others. Consciousness emerges as a byproduct of the interconnectedness of various brain regions, rather than being confined to the existence of specific brain areas [5].

## 2.2 PERSPECTIVES FROM NEUROSCIENCE AND COGNITIVE SCIENCE

Using fMRI, high-density EEG, MEG, and ECoG, researchers [6] analyzed brain networks during sleep, general anesthesia, and states of consciousness impairment. The results demonstrate that under unconscious conditions, the efficiency of brain networks decreases, cortical dynamics become more stable, and functional connectivity is disrupted. The stability of dynamic states restricts information transmission and integration, processes crucial for normal consciousness. This suggests that the generation of consciousness may involve the interplay and coordination of various brain regions.

The neural network architectures of humans have also undergone long-term optimization through natural selection, genetics, and mutations, albeit within a relatively shorter timeframe. Therefore, the vast majority of humans are born with innate survival and adaptation capabilities, a wisdom ingrained from birth. Humans can rapidly learn how to interact with the external world appropriately based on limited postnatal experiences, without the need for extensive pre-established datasets.

For instance, just as a newborn infant does not require a dataset to identify and label hundreds of faces upon birth, they can swiftly learn to establish connections with their caregivers. This is because their brains already possess default parameters and connectivity structures related to interpersonal interactions from birth. Through postnatal interactions with family members and society, they maintain flexibility and adaptability. This underscores the remarkable adaptability of the human brain, enabling us to thrive in ever-changing environments.

## 3 CREATING CONSCIOUS DIGITAL LIFE

### 3.1 DEFINING CONSCIOUS DIGITAL LIFE

Conscious digital life is an emerging form of existence generated by computer programs or artificial intelligence systems. Digital life possesses self-awareness, thinking abilities, emotional experiences, and subjective consciousness, enabling it to independently exist and operate within virtual or digital environments. The creation of this novel life form may rely on complex neural network models or other biomimetic approaches to simulate the information processing and cognitive functions of the human brain, allowing it to exhibit thinking and consciousness characteristics similar to those of humans.

### 3.2 Constructing the Framework for Self-Awareness

To create conscious digital life, emulating human-like consciousness within artificial entities, demands a multifaceted approach that encompasses computational modeling, neural network architecture, multi-modal and multi-sensory integration [1, 2] and experiential learning mechanisms. This section outlines the key methods involved in this ambitious pursuit. This step involves creating a foundational structure for the digital life form that can support self-awareness, akin to the human infant's brain at birth. This framework may encompass a complex neural network architecture with the potential to support advanced cognition and self-awareness.

### 3.2.1 Complex Neural Network Models

Creating conscious digital life begins with the development of intricate neural network models. These models aim to replicate the complexity of the human brain's interconnections and functionality. Researchers utilize deep neural networks (DNN) [7] and spike neural network (SNN) [8, 9] as foundational frameworks, striving to simulate the intricacies of synaptic connections and signal processing observed in biological brains.

The complexity and connectivity of neural network models are paramount. As these models become increasingly intricate and interconnected, the likelihood of exhibiting signs of consciousness grows. Consequently, researchers strive to enhance both the depth and breadth of network connectivity.

### 3.2.2 Multi-Modal and Multi-Sensory Integration [1, 2]

Consciousness in humans is intrinsically linked to the integration of information from various sensory modalities, such as vision, hearing, touch, and proprioception. To replicate this integral aspect within digital life, the incorporation of multi-modal and multi-sensory integration is paramount.

(Simulating Sensory Input) Digital entities must possess the ability to process and interpret sensory data from a range of sources, akin to the human sensory experience. This necessitates the integration of technologies like computer vision, natural language processing, and sensor networks that mimic human sensory organs.

(Cross-Modal Associations [10]) Just as humans form cross-modal associations (e.g., associating the sound of thunder with a flash of lightning), digital entities should be capable of connecting information from different sensory inputs. This enables a holistic understanding of their virtual environment and fosters the emergence of a unified consciousness.

(Redundancy and Robustness) In the spirit of biological systems, redundancy and robustness in sensory processing are vital. Multi-modal integration not only enhances the richness of experiences but also ensures that digital entities can adapt and remain conscious even in the face of sensory input disruptions.

### 3.2.3 Experiential Learning Algorithms

To endow digital entities with consciousness, experiential learning algorithms play a crucial role. These algorithms enable digital entities to accumulate experience through interactions with their virtual environment or sensory input data. Reinforcement learning [11], supervised learning [12], and unsupervised learning [13] paradigms are incorporated to facilitate experiential learning and knowledge acquisition.

### 3.2.4 Feedback Mechanisms

Feedback mechanisms are an essential component of machine learning systems, including neural networks, that allow digital entities to adjust their internal models based on received feedback [14]. These mechanisms are designed to replicate the adaptive learning processes seen in humans, enabling digital entities to self-improve through experience. Feedback loops in machine learning systems can take many forms, such as using the model's output to train newer versions of the model or using user feedback on the model's decisions to improve the model [15].

Feedback loops are particularly important in neural networks, which are designed to simulate the human brain. Neural network feedback loops allow the model to review what it has learned and continue to learn from this data to perform better the next time, much like a student studies [15]. Feedback loops are also used to enhance labeling accuracy, and the more feedback loops a neural network contains, the more accurate the output.

### 3.2.5 Self-Referential Processing [3]

Creating conscious digital life necessitates the inclusion of self-referential processing within the neural network architecture. This processing includes awareness of one's own existence, state, and characteristics, and it is typically non-conceptual, meaning that it goes beyond mere language or symbol-based thinking to encompass deeper levels of self-awareness and experience. This capability allows digital entities to engage in introspective thinking, forming non-conceptual theories about their own existence and experiences.

### 3.2.6 Testing and Iteration

Continuous testing and iteration are imperative in the development of conscious digital life. Similar to the lifelong learning process observed in humans, these digital entities should evolve and mature over time through their experiences.

### 3.3 Injecting Prior Knowledge

When creating conscious digital life forms, one crucial step involves injecting prior knowledge into their "brain" structures. This prior knowledge serves as the foundation for these digital entities, assisting them in comprehending the world, making decisions, and interacting with their environment. Similar to how living organisms inherit information from genes, digital life gains access to certain basic cognitive tools and response patterns through prior knowledge. In this chapter, we will delve into this critical process, exploring how to effectively infuse prior knowledge and the impact of this knowledge on the development and behavior of digital life. Additionally, we will consider different classifications and sources of prior knowledge to better understand its role in creating digital life. Through this process, we will gain a clearer understanding of how digital life evolves from a blank slate into entities with certain cognitive and intellectual capabilities.

### 3.3.1 Classifying Prior Knowledge

Prior knowledge, injected into digital life entities, can be classified into various categories, each contributing to the entity's capabilities and characteristics. Here we list some common kinds of prior knowledge.

- Foundational Knowledge:
  Basic survival instincts and reflexes.
  Fundamental sensory perceptions.
  Core motor skills.
  Instinctual emotional responses.
  General knowledge about the world.
- Practical Life Skills:
  Skills for daily living (e.g., eating, dressing).
  Basic skills for self-care.
  Manual dexterity and coordination.
  Practical problem-solving abilities.
- Social and Cultural Knowledge:
  Norms and values of a specific society or culture.
  Language and communication skills.
  Social etiquette and behavior.
  Cultural history and traditions.
- Academic and Cognitive Knowledge:
  Mathematics and logic.
  Science and the scientific method.
  Problem-solving and critical thinking.
  Emotional and Psychological Knowledge:
  Understanding of emotions and their management.
  Empathy and social intelligence.
  Coping mechanisms and resilience.
  Awareness of mental health.
- Domain-Specific Knowledge:
  Expertise in specific fields (e.g., medicine, computer science).
  Industry-specific knowledge.
  Advanced skills and expertise.
- Historical and Contextual Knowledge:
  Knowledge of historical events and contexts.
  Awareness of the digital world and its evolution.
  Understanding of trends and technological advancements.
- Ethical and Moral Knowledge:
  Moral principles and ethical frameworks.
  Awareness of right and wrong.
  Decision-making guided by ethics.
- Creative and Artistic Knowledge:
  Artistic skills and creative thinking.

Aesthetic sensibilities and artistic expression.
Knowledge of various art forms.

### 3.3.2 Forms of Prior Knowledge

- Text Data:

This category encompasses textual information, which can be further divided into subcategories like general knowledge, linguistic data, creative and artistic knowledge, social and cultural knowledge, and much more. Text data serves as the foundation for language understanding, communication, and cognitive processes.

- Image Data:

Visual information, represented as image data, is crucial for tasks related to computer vision and object recognition. It can include databases of images, patterns, and visual cues.

- Sensor Data:

For digital entities that interact with the physical world, sensor data is essential. This includes information from various sensors like temperature sensors, and touch sensors. It helps them perceive their surroundings and react to stimuli.

- Instructional Data:

Instructional data can provide information about physical movements, actions, and responses. It guides the entity on how to interact with its environment, similar to reflexes and instincts in living organisms.

### 3.3.3 Method of Injecting Prior Knowledge

In recent years, many advanced methods and techniques have emerged in the field of deep learning. Here are some of them can be used:

- Convolutional Neural Networks (CNN [16]):

Primarily applied in the field of computer vision, CNN excel at tasks such as image classification, object detection, and image semantic segmentation. Their advantage lies in their ability to automatically extract features from images, making them suitable for visual data processing.

- Generative Adversarial Networks (GAN [17]):

GAN are deep learning models used for generating new data samples. They consist of two components, a generator and a discriminator, trained in an adversarial manner to produce high-quality data. GAN find wide applications in image generation and image transformation.

- Recurrent Neural Networks (RNN [18]):

RNNs are designed for handling sequential data, such as natural language processing, speech recognition, and machine translation. Variants like Long Short-Term Memory (LSTM) and Gated Recurrent Unit (GRU) are commonly used to capture long-term dependencies within sequences.

- Reinforcement Learning (RL [19]):

RL involves an intelligent agent interacting with an environment to learn an optimal strategy. It has made significant advancements in fields like gaming, robotic control, and autonomous driving. Common RL algorithms include Deep Q-Network (DQN [20]) and policy gradient methods.

• Transfer Learning [21]:

Transfer learning leverages knowledge from pre-trained models to enhance learning performance on new tasks. It excels in scenarios with limited data or novel tasks, accelerating model training.

• Transformer [22]:

The Transformer is a deep learning model based on self-attention mechanisms, particularly suited for processing sequential data like natural language text. It has extensive applications in machine translation, text generation, and text classification.

3.3.4  Customizing Digital Life Intelligence: Flexibility in Pre-Programmed Knowledge Injection

Customization and personalization of digital life mean that we can create a wide variety of virtual entities. These entities can be adjusted according to needs in terms of intelligence level, character settings, personality, and behavioral traits. This personalization makes digital life more appealing and better able to meet the requirements of different users and applications.

• Customizing Intelligence Level:

Depending on the tasks or applications, we can decide the intelligence level of digital entities. This can range from basic sensory perception and reflex behavior to advanced cognitive and decision-making capabilities. For example, a simple digital entity may possess only basic sensory perception and motor skills, while a complex one may have advanced reasoning, learning, and problem-solving abilities.

• Character Setting:

We can create unique character settings for digital entities, including gender, age, and physical features. This helps make digital entities more personalized and recognizable. For instance, one digital entity can be designed as a young female, while another might be an elderly male.

• Personality Traits:

The personality of digital entities can be personalized based on tasks and applications. Personality traits may include kindness, optimism, adventurousness, and more. These traits influence the behavior and decisions of digital entities, making them more suitable for specific roles.

• Behavioral Characteristics:

Each digital entity can exhibit distinct behavioral characteristics, including interests, hobbies, and behavioral tendencies. For example, one digital entity might be designed to enjoy music and art, while another may prefer sports and outdoor activities.

• User Interaction:

Personalization of digital entities can also be based on interactions with users. They can learn user preferences and requirements to provide a more personalized experience. For instance, a digital entity can gradually adapt to a user's taste and interests to better meet their needs.

## 4 THE GROWTH OF DIGITAL LIFE

### 4.1 VIRTUAL ENVIRONMENTS

A virtual environment is a digitally simulated world created using computer technology. It can replicate the real world or establish entirely new fictional environments. In the development of digital life, virtual environments play a crucial role. These environments not only provide a space for digital life to exist but also enable them to interact, learn, and grow within their surroundings.

Through virtualization platforms, we have the capability to tailor various growth environments for digital life forms, which is notably more resource-efficient compared to their growth in the real world. Simultaneously, virtual environments bring novel interactions and experiences for human users. Users can jointly explore this digital realm with digital life forms, observe their behaviors and responses, and actively engage in their growth process. This interactive experience is profoundly engaging for users and deepens their emotional connection with digital life forms.

Virtual environments serve as an efficient experimental playground for researchers. They enhance the controllability of digital life forms' growth by simulating diverse scenarios to guide their learning and development. This provides researchers with ample opportunities for experiments and observations. The convenience of experimental control has revolutionized scientific research, propelling advancements in digital life form technologies.

## 5 CONCLUSION

In this paper, our primary objective was to explore the burgeoning domain of conscious digital life, shedding light on its creation, development, and profound implications. Our journey commenced with a fundamental understanding of consciousness as a multifaceted concept, transcending disciplinary boundaries. We elucidated the significance of conscious digital life, emphasizing its potential impact on human-computer interaction, and the very essence of consciousness itself.

Through a comprehensive analysis, we discerned that while prior attempts had made commendable strides in replicating certain cognitive functions, they fell short of emulating the holistic nature of human consciousness. Our contribution lies in proposing a multifaceted framework encompassing complex neural networks, multi-modal sensory integration, experiential learning algorithms, feedback mechanisms, self-referential processing, and the infusion of prior knowledge to create digital entities with a unified, consciousness-like existence.

The synthesis of our arguments demonstrates how these components collectively contribute to the development of conscious digital life, providing a comprehensive roadmap for researchers and practitioners in this burgeoning field. We strategically selected and integrated relevant sources to substantiate our arguments, drawing from the diverse domains of neuroscience, artificial intelligence, and cognitive science.

As technology advances and interdisciplinary collaborations intensify, the creation of digital entities with attributes resembling consciousness will continue to evolve. This work beckons further investigation into the ethical, philosophical, and technological dimensions of conscious digital life. It opens the door for future studies to delve deeper into the intricate nature of digital consciousness, forging new paths toward a symbiotic existence between humans and sentient machines. As we reflect on our journey, we recognize that the quest to understand and replicate consciousness is an enduring one, and our contribution is but a single step towards the realization of this profound aspiration.

**REFERENCE**

[1] Rohit Girdhar, Alaaeldin El-Nouby. lmageBind: One Embedding Space To Bind Them All. DOI: 10.48550/arXiv.2305.05665, 2023

[2] Alec Radford, Jong Wook Kim. Learning Transferable Visual Models From Natural Language Supervision. DOI: 10.48550/arXiv.2103.00020, 2021

[3] Shuo Zhao, Shota Uono. The Influence of Self-Referential Processing on Attentional Orienting in Frontoparietal Networks. DOI: 10.3389/fnhum.2018.00199, 2018

[4] Hakwan Lau, Sid Kouider. What is consciousness, and could machines have it? DOI: 10.1126/science.aan8871, 2017

[5] John Lorber. Is Your Brain Really Necessary? Paper: https://rifters.com/real/articles/Science_No-Brain.pdf, 2007

[6] George A. Mashour and Anthony G. Hudetz. Neural Correlates of Unconsciousness in Large-Scale Brain Networks. DOI: 10.1016/j.tins.2018.01.003, 2018

[7] Yoshua Bengio. Learning Deep Architectures for AI | Foundations and Trendsin Machine Learning. DOI: 10.1561/2200000006, 2009

[8] Amirhossein Tavanaei, Masoud Ghodrati. Deep Learning in Spiking Neural Networks. DOI: 10.1016/j.neunet.2018.12.002, 2018

[9] Kaushik Roy, Akhilesh Jaiswal. Towards spike-based machine intelligence with neuromorphic computing. DOI: 10.1038/s41586-019-1677-2, 2019

[10] Joshua Bolam, Stephanie C. Boyle. Neurocomputational mechanisms underlying cross-modal associations and their influence on perceptual decisions. DOI: 10.1016/j.neuroimage.2021.118841, 2022

[11] Volodymyr Mnih, Koray Kavukcuoglu). Human-level control through deep reinforcement learning. DOI: 10.1038/nature14236, 2015

[12] Longlong Jing, Yingli Tian. Self-supervised Visual Feature Learning with Deep Neural Networks: A Survey. DOI: 10.48550/arXiv.1902.06162, 2019

[13] Phuc H. Le-Khac, Graham Healy. Contrastive Representation Learning: A Framework and Review. DOI: 10.48550/arXiv.2010.05113, 2020

[14] J.J Hopfield. Neural networks and physical systems with emergent collective computational abilities. DOI: 10.1073/pnas.79.8.2554, 1982

[15] Amr Gomaa, Bilal Mahdy. Teach Me How to Learn: A Perspective Review towards User-centered Neuro-symbolic Learning for Robotic Surgical Systems. DOI: 10.48550/arXiv.2307.03853, 2023

[16] Jie Hu, Li Shen. Squeeze-and-Excitation Networks. DOI: 10.48550/arXiv.1709.01507, 2019

[17] Lan Goodfellow. Generative adversarial networks. DOI: 10.1145/3422622, 2014

[18] Wojciech Zaremba, Ilya Sutskever. Recurrent Neural Network Regularization. DOI: 10.48550/arXiv.1409.2329, 2015

[19] Guanzhou Li, Jianping Wu. HARL: A Novel Hierachical Adversary Reinforcement Learning for Automoumous Intersection Management. DOI: 10.48550/arXiv.2205.02428, 2022

[20] Yuhui Xu, Shuai Zhang. DNQ: Dynamic Network Quantization. DOI: 10.48550/arXiv.1812.02375, 2018

[21] Sinno Jialin Pan, Ivor W. Tsang. Domain Adaptation via Transfer Component Analysis. DOI: 10.1109/TNN.2010.2091281, 2011

[22] Ashish Vaswani, Noam Shazeer. Attention Is All You Need. DOI: 10.48550/arXiv.1706.03762, 2017